\documentclass[amssymb,superscriptaddress]{revtex4}
\usepackage[all,cmtip]{xy}
\usepackage{amsmath,amsfonts,amssymb,amstext}
\usepackage{bbm}
\usepackage{xcolor}
\usepackage{graphicx}
\usepackage{epic,eepic}
\usepackage{pstricks,multido}
\usepackage{pstricks-add}
\font\sqi=cmssq8
\parskip=\medskipamount

\newcommand{\ba}{\begin{array}}
\newcommand{\ea}{\end{array}}
\newcommand{\be}{\begin{equation}}
\newcommand{\ee}{\end{equation}}
\newcommand{\beqs}{\begin{eqnarray}}
\newcommand{\eeqs}{\end{eqnarray}}
\newcommand{\bea}{\begin{eqnarray}}
\newcommand{\eea}{\end{eqnarray}}

\def\DR{\rm I\kern-1.45pt\rm R}
\def\DC{\kern2pt {\hbox{\sqi I}}\kern-4.2pt\rm C}
\def\iprime{\mathcal{I}}
\def\ical{\mathcal{I}}
\def\sfrac#1#2{{\textstyle\frac{#1}{#2}}}

\def\tr#1{\left(#1\right)}

\newcounter{count}
\newpsobject{showgrid}{psgrid}
    {gridlabels=4pt,griddots=0,gridwidth=0.1pt,gridcolor=yellow,subgriddiv=5,subgridwidth=0.05pt,subgridcolor=yellow}
\def\nodeline#1{
\multido{\inode=1+1,\nx=-0.3+0.3}{3}{\psdot[linecolor=blue](\nx,0)\pnode(\nx,0){#1\inode}}
\psellipse[linecolor=lightgray](0,0)(0.5,0.15)
}
\def\multiplet#1(#2,#3)#4{
    \multido{\inode=1+1,\nd=#2+0.3}{#1} {\psdot[linecolor=blue](\nd,#3)\pnode(\nd,#3){#4\inode}}
    \def\ellcen{!0.3\space #1\space 1 sub mul 2 div #2 add #3}
    \def\ellaxes{!0.3\space #1\space 1 sub mul 2 div 0.2 add 0.2}
    \psellipse[linecolor=lightgray](\ellcen)(\ellaxes)
}
\def\multipletx#1(#2,#3)#4{
    \multido{\inode=1+1,\nd=#2+0.3}{#1} {\psdot[linecolor=blue](\nd,#3)\pnode(\nd,#3){#4\inode}}
    \def\ellcen{!0.3\space #1\space 1 sub mul 2 div #2 add #3}
    \def\ellaxes{!0.3\space #1\space 1 sub mul 2 div 0.2 add 0.2}
    \psellipse[linecolor=lightgray](\ellcen)(\ellaxes)
}
\def\conn#1#2{\ncline{#1}{#2}}
\def\spherical(#1,#2){
    \dotnode[linecolor=blue](!#1 #2 1.2 add){up}
    \dotnode[linecolor=blue](#1,#2){down}
    \ncarc[arcangle=25]{down}{up}
    \ncarc[arcangle=-25]{down}{up}
}

\def\red{\text{red}}
\def\spin#1#2{\psi_{#1 #2}}

\begin{document}
\begin{flushright}
ITP--UH--09/11
\end{flushright}

\title{The spherical sector of the Calogero model as a reduced matrix model}

\author{Tigran Hakobyan}
\email{hakob@yerphi.am}
\affiliation{Yerevan State University, 1 Alex Manoogian, 0025 Yerevan, Armenia}
\affiliation{Yerevan Physics Institute, 2 Alikhanyan Br., 0036 Yerevan, Armenia}
\author{Olaf Lechtenfeld}
\email{lechtenf@itp.uni-hannover.de}
\affiliation{ Leibniz Universit\"at Hannover, Institut f\"ur
Theoretische Physik, Appelstr. 2, D-30167 Hannover Germany}
\author{Armen Nersessian}
\email{arnerses@ysu.am}
\affiliation{Yerevan State University, 1 Alex Manoogian, 0025 Yerevan, Armenia}

\begin{abstract}
\noindent\\
We investigate the matrix-model origin of the spherical sector of the
rational Calogero model and its constants of motion. We develop
a diagrammatic technique which allows us to find explicit
expressions of the constants of motion and calculate their Poisson brackets.
In this way we obtain all functionally independent constants of motion
to any given order in the momenta.
Our technique is related to the valence-bond basis for singlet states.
\end{abstract}
\maketitle

\section{Introduction and summary}
\noindent
One of the best known multi-particle integrable systems is the Calogero model
\be
\label{Calogero}
H=\frac{1}{2}\sum_{i=1}^N p_i^2 + \sum_{i<j}\frac{g^2}{(q_i-q_j)^2},\qquad \{ p_i, q_j\}=\delta_{ij}.
\ee
Being
introduced four decades ago \cite{calogero69}, it continues to
attract much interest due to its rich internal structure and
numerous applications. So far, various integrable
extensions have been constructed and studied,
in particular, for the  trigonometric potentials \cite{trig-Cal},
for particles with spins \cite{spin-Cal},
for supersymmetric systems \cite{super-Cal}, and for other Lie algebras \cite{algebra}.
An important feature of all rational Calogero models is the dynamical conformal
symmetry   $so(1,2)\equiv sl(2,R)$, defined  by  the Hamiltonian (\ref{Calogero})
together  with the dilatation $D=\sum_i p_iq_i$ and conformal boost $K= \sum_i q^2_i/2$
generators,
\be
\label{so12}
 \{ {H} , {D}\}= 2 {H}, \quad
\{ {K} , {D}\}=-2 {K}, \quad
\{ {H} , {K}\}=   {D}.
\ee
Due to this symmetry  one can  give an elegant explanation
of the superintegrability property of the conformal invariant
integrable systems \cite{hlkn} (initially observed by Wojcechowski in Calogero
model \cite{woj83}). The ``radial" and ``spherical" parts of rational
Calogero models can be separated in the Hamiltonian
\be
H=\frac{p^2_r}{2}+\frac{{\cal I}(u)}{r^2},\qquad  r\equiv \sqrt{2K},\quad
 p_r\equiv \frac{D}{\sqrt{2K}}:\qquad\{p_r, r\}=1,\quad \{p_r, u^\alpha\}=\{r, u^\alpha\}=0,
\label{sC}
\ee
with the "spherical part" corresponding to the Casimir element of the conformal algebra.
Hence, the whole information about the conformal mechanics is encoded in its ``spherical part",
 given by the Hamiltonian system
\be
{\cal I}(u) \qquad\textrm{with}\qquad
\{u^\alpha, u^\beta\}=\omega^{\alpha\beta}(u).
\ee
This system  is of its own interest since it
 describes a multi-center generalization of the
$(N-1)$-dimensional Higgs oscillator \cite{cuboct}.
In the quantum case and for special discrete values of the
coupling constant it can be mapped to free-particle systems on
the sphere \cite{feigin}.
However, the connection between the
constants of motion of the initial conformal mechanics and its
spherical sector is highly complicated \cite{hlkn,hlns}.
In particular, it is unclear up to now how to construct the
Liouville constants of motion of the spherical sector from the ones
of the full conformal mechanics.

On the other hand, the rational Calogero model can be easily
constructed from the free Hermitian matrix model via a Hamiltonian
reduction~\cite{perelomov}. In this way, we get a transparent
explanation of its integrability property and the Lax pair
formulation. Hence, it is natural to try to explore the matrix
origin of the spherical sector of the Calogero model in order
to find the matrix-model origin of its constants of motion.
In that case we shall immediately get the constants of motion of the spin-Calogero
model as well. One may expect that the matrix model formulation of the spherical sector of
the Calogero model can simplify the study of its constants of motion.
 Moreover, such a formulation,  being purely algebraic, might establish
new relations between the spherical sector of the Calogero
model and other algebraic integrable systems, for instance lattice spin systems.

The investigation of the spherical sector of the rational Calogero
model and of its constants of motion  at the matrix-model level is
the goal of the present paper.
First we recall the formulation of the Liouville
integrals of the initial Calogero model at the
matrix-model level in terms of $U(N)$-invariant polynomials (and of
$SU(N)$-invariant polynomials for the Calogero model with the center of mass excluded)
corresponding to the highest states of the conformal algebra.
Then we observe that the constants of motion of the spherical system
are described by $SU(N)\times SL(2,R)$ singlets. This allows us to reduce the study of the algebra
of invariants of the spherical sector to a purely algebraic computation
of $SU(N)$ invariant tensors.
To simplify the calculations, we develop an appropriate diagrammatic
technique illustrated by numerous examples.
We present explicit expressions for all functionally independent constants of motion
up to sixth order in momenta as well as recover the results obtained  in \cite{hlns}
by the use of standard methods.
Finally, we  establish a relation of the developed diagrammatic technique
with the valence-bond basis introduced by Temperley and Lieb \cite{TL71}.

The paper  is arranged as follows.
In Section~2 we give a brief description of the matrix-model formulation
of the rational Calogero model, including the description of the
reduction procedure and the exclusion of the center of mass at the matrix-model frame.
Then  we develop a similar formulation for the spherical sector of the Calogero model.
In Section~3 we develop the diagrammatic technique for the formulation of the constants of motion
of the spherical sector of the Calogero model and find by its use all functionally
independent constants of motion up to sixth order in momenta.
In Section~4, considering the free-particle limit, we rederive,
by the use of our technique, the constants of motion obtained
in \cite{hlns} by standard methods.
In Section~5 we establish a  correspondence between
our technique and the valence-bond basis developed by Temperley and Lieb.
An interesting future task concerns the relation of the symmetries of the
spherical sector of the Calogero model with $W$- and Hecke algebras.

\section{Matrix-model formulation}
\noindent
We recall that the Calogero model \eqref{Calogero}
has $N$ Liouville constants of motion
\cite{moser,polychronakos,gauge}
\be
\label{Liouville}
I_k=\tr{L^k}:=\text{tr} \, L^k, \qquad 1\le k \le N,
\ee
given in terms of the Lax matrix
\be
\label{Lax}
L_{jk}=\delta_{jk}p_k+(1{-}\delta_{jk})\frac{ig}{q_j-q_k}.
\ee
For convenience, hereafter the trace of a matrix is denoted by round brackets.
The Calogero model can be obtained from the free Hermitian matrix model
\be
\label{matrix}
H=\frac12 \tr{P^2}, \qquad P^+=P, \qquad Q^+=Q,
\qquad \{P_{ij},Q_{j'i'}\}=\delta_{ii'}\delta_{jj'}
\ee
by the reduction corresponding to the $SU(N)$ group action
\be
\label{shift}
P\to UPU^+, \qquad Q\to UQU^+,
\ee
which also preserves the canonical brackets.
The related conserved current is given by the traceless Hermitian matrix \cite{polychronakos}
\be
\label{Jmatrix}
J=i[Q,P].
\ee
Using this symmetry, one can diagonalize  the coordinate matrix
\be
\label{redQ}
Q_{jk}\to q_j\delta_{jk}.
\ee
Then, according to the relation \eqref{Jmatrix}, the diagonal matrix elements of $J$
vanish while the off-diagonal ones define
the related elements of the reduced matrix $P$:
$$
P_{jk}\to -i\frac{J_{jk}}{q_j-q_k}, \qquad j\ne k.
$$
The diagonal elements $p_j$ of the reduced matrix remain independent and are
conjugate to the respective coordinates, i.e.~$\{p_j,q_k\}=\delta_{jk}$.
The simplest choice
$J_{jk}=-g$ for all $j\ne k$
recovers the Lax matrix \eqref{Lax} of the usual Calogero model:
\be
\label{redP}
P_{jk}\to L_{jk}.
\ee

After the gauge fixing, the commutator \eqref{Jmatrix} reduces to
\be
\label{redPQ}
 [Q,P]_{jk}=ig (\delta_{jk}-u_ju_k), \qquad \text{where $u_i=1$}.
\ee
Let us choose an orthogonal basis $\{T_a\}$, $a=0,1,\ldots,N^2{-}1$,
for $u(N)$ with
\be
\label{Ta}
\tr{T_aT_b}=\delta_{ab},
\qquad
[T_a,T_b]=\sum_c if_{abc}T_c
\ee
and real antisymmetric structure constants $f_{abc}$ and expand
the Hermitian  matrices
\be
\label{QaPa}
Q=\sum_aQ_aT_a,
\qquad
P=\sum_aP_aT_a.
\ee
According to the decomposition $U(N)=U(1)\times SU(N)$, we take the identity matrix
as a $U(1)$ generator $T_0$, and the remaining $T_a$ are given by traceless
Hermitian matrices.
The standard form of the basis is given by \eqref{cartan1}, \eqref{cartan2}, and
\eqref{borel} in the  Appendix.

The coefficients $P_a$ and $Q_a$ form pairs of conjugate momenta and coordinates \cite{poly99,polychronakos}
$$
\{P_a,Q_b\}=\delta_{ab},
$$
as it follows from the Poisson brackets in \eqref{matrix} and the completeness relation
among the $U(N)$ generators \eqref{comp1}.
We define the angular momentum tensor $M$ for the matrix model by
\be
\label{Mab}
M_{ab}=P_aQ_b-P_bQ_a.
\ee
The angular momentum tensor components $M_{ab}$ form an $SO(N^2)$ algebra
\be
\label{soN}
\{M_{ab},M_{a'b'}\}=\delta_{ab'}M_{a'b}+\delta_{ba'}M_{ab'}-\delta_{aa'}M_{bb'}-\delta_{bb'}M_{aa'}.
\ee
The matrix form of angular momentum tensor is
\be
\label{momentum}
M= P\wedge Q\equiv P\otimes Q-Q\otimes P=\sum_{a,b}M_{ab}\,T_a\otimes T_b.
\ee
The components $J_c=\tr{J T_c}$ of the $SU(N)$ conserved current  \eqref{Jmatrix} are expressed in terms of
the angular momentum components as
\be
\label{Jc}
J_c=-\frac{i}{2}\sum_{a,b}f_{abc}M_{ab}.
\ee
They obey, of course, the $su(N)$ algebra
\be
\label{suN}
\{J_a,J_b\}=i\sum_c f_{abc}J_c,
\ee
as can be verified independently using \eqref{soN}.

There is an additional $U(1)$ symmetry of the matrix system given by
the translations
\be
\label{translation}
Q\to Q+\epsilon \mathbf{1}, \qquad P\to P.
\ee
The related conserved  current is (see \eqref{a0} in the Appendix)
\be
\label{P0}
\tr{P}\sim P_0
\ee
The $SU(N)$ shifts preserve it, so it is in involution with the
$SU(N)$ currents $J_a$, which can be verified also using the
relations \eqref{matrix} and \eqref{Jmatrix}.
Together these currents generate the $U(1)\times SU(N)=U(N)$ group.
The $U(1)$ reduction eliminates of the center-of-mass
momenta and coordinates of the Calogero system,
$$
\tr{P}=\sum_i p_i=0,
\qquad
\tr{Q}=\sum_i q_i=0,
$$

The action
 of the matrix model, $S=\sfrac12\int \text{tr}\dot{Q}^2dt$, remains invariant
under the conformal transformations forming the $sl(2,R)$ algebra~\eqref{so12}
generated by the quantities
\be
\label{gen-sl2}
D=\tr{PQ}=\sum_a Q_aP_a, \qquad
K=\frac12\tr{Q^2}=\frac12\sum_a Q_a^2, \qquad
H=\frac12\tr{P^2}=\frac12\sum_a P_a^2.
\ee
The standard basis of this algebra has the form
\be
\label{sl2-inv}
J_1=H+K, \qquad J_2=D, \qquad J_3=H-K,
\qquad
\{J_\alpha,J_\beta\}=-2\epsilon_{\alpha\beta\gamma}J^\gamma,
\ee
where the indices are raised by the conformal metric $\text{diag}(1,-1,-1)$.
The conformal algebra is in involution with the angular momentum tensor
\be
\label{M-singlet}
\{M_{ab},sl(2,R)\}=0
\ee
and, hence, with the $su(N)$ algebra \eqref{Jc}, \eqref{suN}. The last fact follows also
from the invariance of the traces \eqref{gen-sl2} under the gauge transformations
\eqref{shift}.

The  Casimir
element of the algebra \eqref{so12} looks as follows,
\be \label{casimir}
\iprime=\sum_\alpha J_\alpha
J^\alpha=4{KH}-{D}^2=\tr{P^2}\tr{Q^2}-\tr{PQ}^2,
 \ee
 and  defines the Hamiltonian of the matrix-model  origin of the
spherical part of the Calogero model \eqref{sC},
which can be considered   as a separate system  and hereafter will be called ``spherical mechanics".
It can also be
expressed  in terms of the  angular momentum \eqref{Mab},
\be
\ical=\frac12(\text{tr}\otimes\text{tr})M^2=\sum_{a<b}M_{ab}^2.
\ee
We note that the Casimir  maps  the Liouville
constants of motion \eqref{Liouville} to $N-1$ additional ones \cite{hlkn,hlns},
which are responsible for the superintegrability of  the Calogero model
\cite{woj83}:
\be \label{Gk} G_k=\{\ical,I_k\}, \qquad
k=1,3,4,\dots,N .\ee

Note that the conformal generators \eqref{gen-sl2} are composed from independent parts, each
specified by one coordinate and momentum component. However, as was mentioned above,
the $SU(N)$ reduction to the Calogero model \eqref{redQ}, \eqref{redP} mixes together
all components apart from the first one, which corresponds to the center of mass.
Therefore, we have a well defined decomposition of mutually involutive conformal algebras,
\be
J_\alpha=J_\alpha^\red+J_\alpha^0,
\ee
where $J_\alpha^0$ is defined by the $a=0$ term in the sums \eqref{gen-sl2}
$$
D_0=P_0Q_0, \qquad
K_0=\frac12 Q_0^2, \qquad
H_0=\frac12 P_0^2,
$$
while
$J_\alpha^\red$ is determined by the others. For the Casimir element, we have
\be
\label{ical-red}
\ical=\ical^\red+2J_\alpha^\red J^{0\,\alpha}.
\ee
This is the relation between the spherical systems with and without center of mass.

\section{The constants of motion  of the spherical mechanics}
\noindent
Before discussing the constants  of motion of the spherical mechanics, we rewrite
the Liouville constants of motion of the Calogero system \eqref{Liouville}
in terms of matrix model generators,
\be
\label{Ik}
I_k=\tr{P^k}=\sum_{a_1,\dots,a_k}d_{a_1\dots a_k}P_{a_1}\dots P_{a_k},
\qquad
0\le a_i \le N^2-1,
\ee
where the coefficients $d_{a_1\dots a_k}$ are $U(N)$ invariant tensors  defined by the expressions
\be
\label{inv-tensor}
d_{a_1\dots a_k}=\tr{T_{a_1}\dots T_{a_k}}.
\ee
In particular,
\be
\label{dab}
d_a=\delta_{a0}, \qquad d_{ab}=\delta_{ab}.
\ee
The first two constants of motion  are proportional to the total momentum and Hamiltonian
of the system, respectively.
Setting $P_0=0$ in \eqref{Ik}, we obtain the constants of motion  for the system
with excluded center of mass,
\be
I_k^\red=\sum_{b_1,\dots,b_k}d_{b_1\dots b_k}P_{b_1}\dots P_{b_k}, \qquad
1\le b_i \le N^2-1.
\ee
Since the first constant of motion vanishes due to \eqref{dab}, only $N-1$ independent
Liouville integrals remain.
Then the relation between the constants  of motion of the Calogero system with and without
mass center reads
\be
\label{int-un-sun}
I_k\to\sum_{i=0}^k  N^\frac{i-k}{2}\binom{k}{i}P_0^{k-i} I_i^\red,
\ee
where we set $I^\red_0=1$ and $I^\red_1=0$.
It is a consequence of the relation between the invariant tensors of the $SU(N)$ and $U(N)$ groups,
\be
\label{d-un-sun}
d_{a_1\dots a_k}=N^{-\frac{k-k'}{2}}d_{a_{r_1}\dots a_{r_{k'}}}
\ee
 where $r_1,\dots,r_{k'}$ are
the positions of the indices with nonzero values taken in ascending order, i.e.
$a_{r_i}>0$. The equation \eqref{d-un-sun} follows from \eqref{cartan1}, \eqref{cartan2}
and \eqref{inv-tensor}.
A similar relation can be derived between the additional integrals $G_k$ of both systems
using their expression \eqref{Gk} and \eqref{ical-red}.

The constants of motion of the spherical mechanics \eqref{casimir} have to be in
involution with the whole conformal algebra \eqref{so12} since they
are expressed in terms of the angular coordinates and momenta, while
the conformal algebra generators depend on the radial coordinate and
momentum only. Therefore, they are  $sl(2,R)$ singlets.
On the other hand, the integrals must be also $SU(N)$ scalars
in order to assure a valid reduction
map. So, \emph{the algebra of integrals of  the spherical
mechanics is formed by $SU(N)\times SL(2,R)$ singlets.}
In this section, we construct them from $SL(2,R)$ invariants by combining them in an
appropriate way in order to obtain an $SU(N)$ invariant. This approach
provides the integrals of the spherical Hamiltonian with
a simple graphical picture.

As was mentioned above, any $SL(2,R)$
invariant can be expressed in terms of angular momentum tensor
components $M_{ab}$ with indices belonging to the adjoint
representation of $SU(N)$. At the same time, an $SU(N)$ invariant can be
constructed by contraction of the monomials $M_{a_1b_1}\dots M_{a_kb_k} $
with a number of invariant tensors \eqref{inv-tensor}. This
observable will be a  polynomial constant of motion of
$\ical$ of $k$th order both in $M_{ab}$ and momenta $P_a$. It can be presented
in graphical form by drawing the angular momentum tensor as a vector with the endpoints
endowed with the corresponding indices. The aligned endpoints inside a cycle
mean the contraction of the related indices with the invariant tensor as is
shown in Fig.~\ref{fig:Mab}.
\begin{figure}
\psset{arrows=->,dotsize=2pt}
\psset{linecolor=brown!70!black}
\begin{center}
\begin{pspicture}(0,0)(3,0.5)
\put(0,0){$M_{ab}=$}
\pnode(1.5,2pt){a}
\pnode(2.8,2pt){b}
\ncline{a}{b}
\uput[u](1.5,1pt){\small $a$}
\uput[u](2.8,1pt){\small $b$}
\end{pspicture}
\begin{pspicture}(-2,0)(2,0.5)
\put(0,0){$d_{abcde}=$}
\multido{\in=1+1,\nx=2+0.4}{5}{
\psdots[linecolor=blue!80!black](\nx,2pt)
\setcounter{count}{\the\multidocount}
\uput[u](\nx,2pt){\small $\alph{count}$}
}
\psellipse[linecolor=lightgray](2.8,2pt)(1.2,0.15)
\end{pspicture}
\end{center}
\caption{\label{fig:Mab}
Graphical representations of the angular momentum and
invariant tensors.}
\end{figure}
The entire diagram consists of vectors with endpoints distributed along legs of such type.
Among these diagrams, some are expressed in terms of others or vanish.
In particular, the quadratic bond-crossing relations among the components of the momentum tensor,
\be
\label{crossing}
M_{a'b}M_{ab'}=M_{ab}M_{a'b'}-M_{aa'}M_{bb'},
\ee
which is a consequence of the definition \eqref{momentum} and
 presented diagrammatically in Fig.~\ref{fig:crossing},
reduces significantly the number of  functionally independent integrals.
\psset{arrows=->,dotsize=2pt}
\psset{linecolor=brown!70!black}
\begin{figure}
%\begin{center}
\begin{pspicture}(10,1.5)
\psline(0,0)(0,1)
\psline(1,0)(1,1)
\psline(3,0)(4,1)
\psline(4,0)(3,1)
\psline(6,1)(7,1)
\psline(6,0)(7,0)
\rput[br](-0.1,0){$a$}
\rput[br](-0.1,1){$b$}
\rput[bl](1.1,0){$a'$}
\rput[bl](1.1,1){$b'$}
\rput[br](2.9,0){$a$}
\rput[br](2.9,1){$b$}
\rput[bl](4.1,0){$a'$}
\rput[bl](4.1,1){$b'$}
\rput[br](5.9,0){$a$}
\rput[br](5.9,1){$b$}
\rput[bl](7.1,0){$a'$}
\rput[bl](7.1,1){$b'$}
\rput(2,0.5){\large$=$}
\rput(5,0.5){\large$+$}
\end{pspicture}
%\end{center}
\caption{\label{fig:crossing}Diagrammatic representation of the crossing relations \eqref{crossing} among $M_{ab}$. }
\end{figure}

It is clear that an invariant corresponding to a disconnected diagram is just the product
of the invariants corresponding to its connected parts.
Although the order of indices inside the leg is relevant up to cyclic permutations, any
permutation changes the observable by integrals of similar type carrying lower orders in the momenta.
This follows from the crossing relations \eqref{crossing}, the antisymmetry of the angular momentum tensor
and the definition of the $SU(N)$ current $J_c$ \eqref{Jc}, which becomes a number upon the reduction:
$$
\sum_{a,b}\tr{\dots[T_a,T_b]\dots}M_{aa'}M_{bb'}
=\frac{i}{2}\sum_{a,b,e}f_{abe}\tr{\dots T_e\dots}M_{ab}M_{a'b'}
=-\frac{1}{2}\sum_{e}J_e d_{\dots e\dots} M_{a'b'}.
$$
Moreover, if two indices of an invariant tensor are attached to the same $M_{ab}$,
the corresponding integral is reduced to a combination of lower-order integrals.
Indeed, for contracted adjacent indices we have
$$
\sum_{a,b}d_{ab\dots}M_{ab}=\sum_{a,b}\tr{T_aT_b\dots}M_{ab}
=\sum_{a,b,c}\frac{i}{2}f_{abc}d_{c\dots}M_{ab} =-\sum_c J_c d_{c\dots}.
$$
\psset{arrows=->,dotsize=2pt}
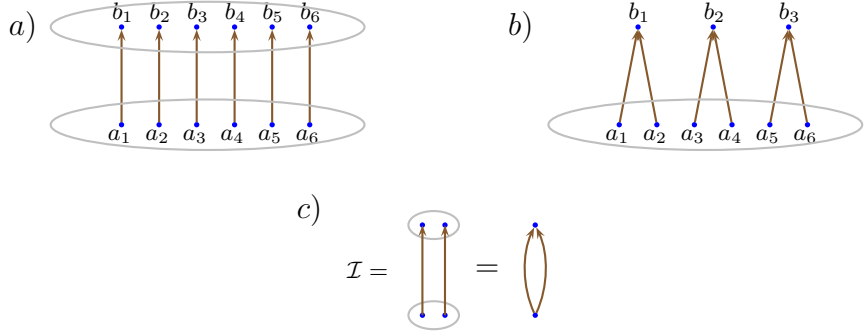
\begin{figure}
\begin{pspicture}(-1.4,-0.6)(4,2.2)
\rput(-1.2,1.3){\large$a)$}
\multiput(0,0)(0.5,0){6}
{
\dotnode[linecolor=blue](0,0){1}
\dotnode[linecolor=blue](0,1.3){2}
\ncline{1}{2}
}
\multido{\nx=1+1}{6}
{
\uput{2pt}[290](!\nx \space 1 sub 0.5 mul 0){\small $a_\nx$}
\uput{2pt}[60](!\nx \space 1 sub 0.5 mul 1.3){\small $b_\nx$}
}
\psellipse[linecolor=lightgray](1.25,0)(2.1,0.35)
\psellipse[linecolor=lightgray](1.25,1.3)(2.1,0.35)
\end{pspicture}
\hspace{1cm}
\begin{pspicture}(-1.4,-0.6)(4,2.2)
\rput(-1.2,1.3){\large$b)$}
\multiput(0,0)(1,0){3}
{
\dotnode[linecolor=blue](0,0){1}
\dotnode[linecolor=blue](0.25,1.3){3}
\dotnode[linecolor=blue](0.5,0){2}
\ncline{1}{3}
\ncline{2}{3}
}
\multido{\nx=1+1}{6}
{
\uput{2pt}[290](!\nx \space 1 sub 0.5 mul 0){\small $a_\nx$}
}
\multido{\nx=1+1}{3}
{
\uput{2pt}[60](!\nx \space 0.75 sub 1.3){\small $b_\nx$}
}
\psellipse[linecolor=lightgray](1.25,0)(2.1,0.35)
\end{pspicture}
\hspace{1cm}
\begin{pspicture}(-1.4,-1.2)(3,1.3)
\rput(-1.2,0.8){\large$c)$}
\rput[l](-0.7,0){$\mathcal{I}=$}
\multiplet{2}(0.3,-0.6){A}
\multiplet{2}(0.3,0.6){B}
\conn{A1}{B1}\conn{A2}{B2}
\rput[l](1,0){\large$=$}
\spherical(1.8,-0.6)
\end{pspicture}
\caption{\label{fig:examples-1} Diagrammatic representations for the integrals $\ical_k$ $(a)$ and $\ical'_k$ $(b)$ and the
spherical Hamiltonian $(c)$.
}
\end{figure}
Therefore, without loss of generality, one  may consider a diagram topologically equivalent to one
with invariant tensors located along a single line or cycle with mutually nonintersecting angular momentum bonds.

\begin{figure}
\begin{pspicture}(-1.6,-1.2)(1.2,1.2)
\rput(-1.4,0.8){\large$a)$}
\multido{\nangle=-45+90,\iangle=45+90}{4}
{
\setcounter{count}{\the\multidocount}
\rput{\nangle}(1;\iangle){\nodeline{\Alph{count}}}
}
\conn{B3}{A1}
\conn{B2}{A2}
\conn{C1}{D3}
\conn{C2}{D2}
\conn{D1}{A3}
\conn{C3}{B1}
\end{pspicture}
\hspace{2cm}
\begin{pspicture}(-2.4,-1.2)(2,1.2)
\rput(-2.2,0.8){\large$c)$}
\rput(-1.4,0){\large$\ical_{4,3}=$}
\multiplet{5}(-0.1,0.6){A}
\multiplet{4}(-0.6,-0.6){B}
\multiplet{3}(1.1,-0.6){C}
\conn{B1}{A1}
\conn{B2}{A2}
\conn{B3}{A3}
\conn{C2}{A4}
\conn{C3}{A5}
\conn{B4}{C1}
\end{pspicture}
\hspace{1cm}
\begin{pspicture}(-2,-1.2)(3.6,1.2)
\rput(-1.6,0.8){\large$d)$}
\rput(-0.8,0){\large$\ical_{4,6}=$}
\multiplet{4}(0,-0.6){A}
\multiplet{6}(1.8,-0.6){B}
\multiplet{8}(0.6,0.6){C}
\conn{A1}{C1}
\conn{A2}{C2}
\conn{A3}{C3}
\conn{A4}{B1}
\conn{B2}{C4}
\conn{B3}{C5}
\conn{B4}{C6}
\conn{B5}{C7}
\conn{B6}{C8}
\end{pspicture}
\begin{pspicture}(-1.6,-1.2)(1.2,1.2)
\rput(-1.4,0.8){\large$b)$}
\multiplet{3}(-0.8,0.6){B}
\multiplet{1}(0.8,0.6){C}
\multiplet{4}(-0.5,-0.6){A}
\conn{A1}{B1}\conn{A2}{B2}\conn{A3}{B3}
\conn{A4}{C1}
\end{pspicture}
\hspace{2cm}
\begin{pspicture}(-2.4,-1.2)(2,1.2)
\rput(-2.2,0.8){\large$e)$} \rput(-1.4,0){\large$\ical'_{4,3}=$}
\multiplet{3}(-0.6,0.6){Aa}
\multiplet{2}(1.4,0.6){Ab}
\multiplet{4}(-0.6,-0.6){B}
\multiplet{3}(1.1,-0.6){C}
\conn{B1}{Aa1} \conn{B2}{Aa2} \conn{B3}{Aa3}
\conn{C2}{Ab1} \conn{C3}{Ab2} \conn{B4}{C1}
\end{pspicture}
\hspace{1cm}
\begin{pspicture}(-2,-1.2)(3.6,1.2)
\rput(-1.6,0.8){\large$f)$} \rput(-0.8,0){\large$\ical'_{4,6}=$}
\multiplet{4}(0,-0.6){A}
\multiplet{6}(1.8,-0.6){B}
\multiplet{3}(0,0.6){Ca}
\multiplet{5}(2.1,0.6){Cb}
\conn{A1}{Ca1} \conn{A2}{Ca2} \conn{A3}{Ca3}
\conn{A4}{B1} \conn{B2}{Cb1} \conn{B3}{Cb2} \conn{B4}{Cb3}
\conn{B5}{Cb4} \conn{B6}{Cb5}
\end{pspicture}
\caption{\label{fig:examples-2} Diagrammatic representations of some spherical invariants. The invariants of type
$\ical_{n,m}$ and $\ical'_{n,m}$ are obtained from the commutators of $\ical_n$ and $\ical_m$ drawn in Fig.~\ref{fig:examples-1}.
}
\end{figure}
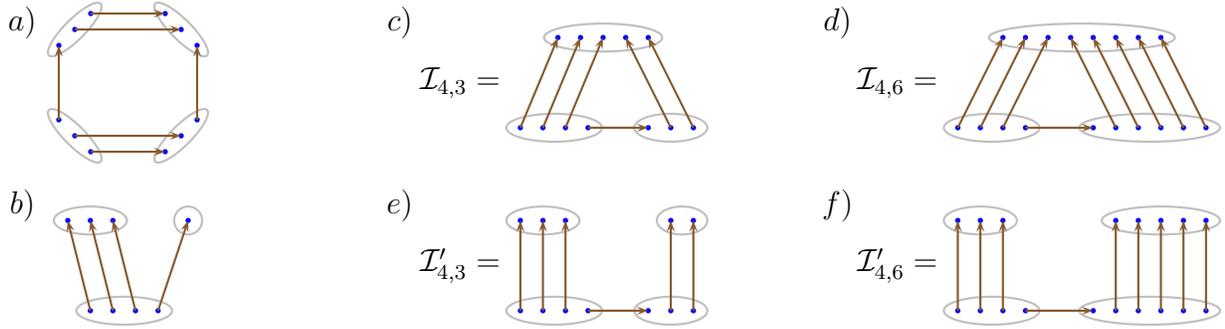

As an example, consider the analogue of the Liouville constants of motion \eqref{Ik} of the original Calogero
Hamiltonian. For the spherical Hamiltonian, one must use the angular momentum instead of the momentum:
\begin{eqnarray}
\label{Ik-sp}
\ical_k=\tr{M^{2k}}:=({\rm tr}\otimes{\rm tr})M^{2k}
=\sum_{a_i,b_i} d_{a_1\dots a_{2k}}d_{b_1\dots b_{2k}}M_{a_1b_1}\dots M_{a_{2k}b_{2k}}.
\end{eqnarray}
Note that due to the antisymmetry of $M_{ab}$, the odd powers vanish.
The related diagram is shown in Fig.~\ref{fig:examples-1}a.
In contrast to their analogue \eqref{Ik}, these spherical
integrals are not in involution. Their bracket equals
\be
\label{Inm}
\{\ical_n,\ical_m\}=4nm\ical_{n,m},
\ee
where $\ical_{n,m}$ is a combined diagram glued in a way presented in
Figs.~\ref{fig:examples-1}b and \ref{fig:examples-1}c. The above relation
is a consequence of \eqref{soN}, the cyclic symmetry of the invariant tensors \eqref{inv-tensor}
and the completeness relation among them,
\be
\label{d-comp1}
\sum_{a=0}^{N^2-1}d_{a_1 \dots a_{n-1}a}d_{a a_n \dots a_{n+m-1}}=d_{a_1a_2 \dots a_{n+m-1}},
\ee
which follows from \eqref{comp1}.
For the system with  reduced center of mass, the last relation
reads, according to \eqref{comp2},
\be
\label{d-comp2}
\sum_{b=1}^{N^2-1}d_{b_1 \dots b_{n-1}b}d_{b b_n \dots b_{n+m-1}}=d_{b_1b_2 \dots b_{n+m-1}}
+\frac1N d_{b_1 \dots b_{n-1}}d_{b_n \dots b_{n+m-1}}.
\ee
As a consequence,  the commutator \eqref{Inm} acquires the following form:
\be
\label{Inm-red}
\{\ical^\red_n,\ical^\red_m\}=4nm\left(\ical^\red_{n,m}-\frac1N\ical^{'\red}_{n,m}\right),
\ee
where $\ical'_{n,m}$ is derived from $I_{n,m}$ by splitting its longest leg
as is shown in Fig.~\ref{fig:examples-2}.

\begin{figure}
\begin{pspicture}(-0.5,-0.4)(5.6,1)
\multiplet{2}(0,-0.2){A}
\dotnode[linecolor=white](-0.3,0.8){B1}
\dotnode[linecolor=white](0.6,0.8){B2}
\rput(-0.2,0.9){$a$}
\rput(0.7,0.9){$b$}
\conn{A1}{B1}
\conn{A2}{B2}
\rput(1.2,0.2){\large$=$}
\dotnode[linecolor=white](1.75,0.8){C1}
\dotnode[linecolor=white](2.55,0.8){C2}
\dotnode[linecolor=blue](2.15,-0.2){D}
\rput(1.75,0.9){$a$}
\rput(2.65,0.9){$b$}
\conn{D}{C1}
\conn{D}{C2}
\uput[r](2.8,0.2){\large$=\;\;\sum_c M_{ca}M_{cb}$}
\end{pspicture}
\hspace{2cm}
\begin{pspicture}(-0.5,-0.4)(2.7,1)
%\showgrid
\multiplet{1}(0,-0.2){A}
\dotnode[linecolor=white](0,0.8){B}
\conn{A1}{B}
\rput(0.2,0.8){$a$}
\rput(0.8,0.2){\large $=$}
\dotnode[linecolor=white](1.4,-0.2){C}
\dotnode[linecolor=white](1.4,0.8){D}
\rput(1.6,0.8){$a$}
\rput(1.6,-0.2){$0$}
\conn{C}{D}
\rput(2.6,0.2){\large $=\;\;M_{0a}$}
\end{pspicture}
\caption{\label{fig:double-dot}A double dot leg can be replaced by a single dot.}
\end{figure}
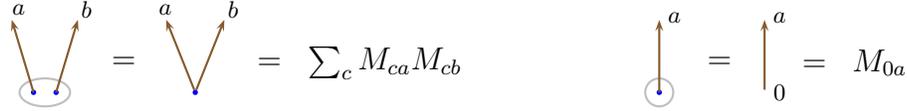

Since $d_{ab}=\delta_{ab}$, a double-dot leg can be replaced by a single dot as
shown on Fig.~\ref{fig:double-dot}. So,
the first integral from the set \eqref{Ik-sp} just coincides with the spherical
Hamiltonian itself: $\ical_1=\ical$ (see Fig.~\ref{fig:examples-2}c).
The single-dot leg exists only in the presence of the center of mass,
because $d_a=\delta_{0a}$ (see the same figure). This property reduces significantly
the number of independent invariants for the spherical mechanics without mass center
and simplifies their classification.

Another family of constants of motion  can be obtained  by contracting
the first indices of two adjacent angular momentum tensors
(via the invariant  $\delta_{aa'}$) and contracting their second indices
with some $2k$-th order invariant tensor:
\be
\label{Ik'}
\ical'_k=\tr{R^k}=\sum_{a_i,b_i,b'_i}d_{b_1b'_1\dots b_k b'_k} M_{a_1b_1}M_{a_1b'_1}\dots M_{a_kb_k}M_{a_kb'_k},
\qquad
R=({\rm tr}\otimes 1) M^2.
\ee
This set has been considered in \cite{woj84} as constants of motion for the system with the Hamiltonian
$HK$. Note that the first constant of motion  from this set also coincides with the spherical
Hamiltonian, $\ical'_1=\ical$.\\

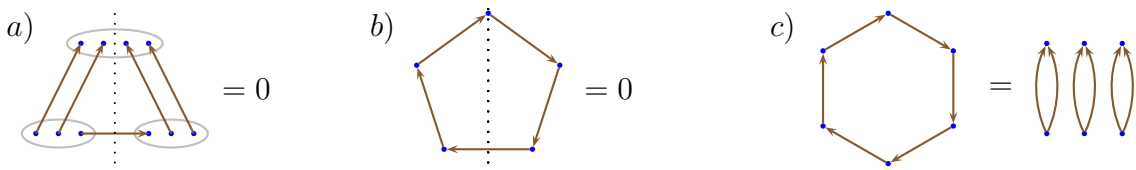
\begin{figure}
\begin{pspicture}(-1,-1.2)(2.6,1.2)
\rput(-0.8,0.8){\large$a)$}
\multiplet{4}(0,0.6){A}
\multiplet{3}(-0.6,-0.6){B}
\multiplet{3}(0.9,-0.6){C}
\conn{B1}{A1}
\conn{B2}{A2}
\conn{C2}{A3}
\conn{C3}{A4}
\conn{B3}{C1}
\psline[linestyle=dotted,linecolor=black]{-}(0.45,-1)(0.45,1)
\rput(2.2,0){\large$=0$}
\end{pspicture}
\hspace{1cm}
\begin{pspicture}(-1.6,-1.2)(2.3,1.2)
\rput(-1.4,0.8){\large$b)$}
\multido{\inode=0+1,\nangle=18+72}{5}{\dotnode[linecolor=blue](1;\nangle){\inode}}
\multido{\inode=0+1,\inext=1+1}{5}{
\ifnum\inode=4 \def\inext{0} \fi
\ncline{\inext}{\inode}
\psline[linestyle=dotted,linecolor=black]{-}(0,-1)(0,1.2)
}
\rput(1.6,0){\large$=0$}
\end{pspicture}
\hspace{1cm}
\begin{pspicture}(-1.8,-1.2)(3.3,1.2)
\rput(-1.4,0.8){\large$c)$}
\multido{\inode=0+1,\nangle=30+60}{6}{\dotnode[linecolor=blue](1;\nangle){\inode}}
\multido{\inode=0+1,\inext=1+1}{6}{
\ifnum\inode=5 \def\inext{0} \fi
\ncline{\inext}{\inode}
}
\rput(1.5,0){\large$=$}
\multiput(2,-0.6)(0.5,0){3}{\spherical(0,0)}
\end{pspicture}
\caption{\label{fig:polygon}
The two first diagrams correspond to vanishing invariants:
they undergo a sign change
under the reflection with respect to the symmetry axis indicated by the dotted line.
The invariants \eqref{polygon} are given by an $n$-sided polygon. }
\end{figure}

More general invariants  may contain nontrivial loops of angular momentum bonds, which include more
invariant tensors.
The number of bonds yields the order in momenta or angular momenta of the related invariant.
In Figs.~\ref{fig:examples-2},  \ref{fig:polygon} and \ref{fig:four-order} some other examples of constants
of motion of the spherical Hamiltonian are presented. Let us write down, for example, the invariant
corresponding to the sixth-order diagram in Fig.~\ref{fig:four-order}c:
$$
\sum_{\text{all indices}} d_{abc}d_{a'b'c'}M_{ad}M_{a'd}M_{bb'}M_{cc'}.
$$

Of course, only a finite number of diagrams are functionally independent. In the presence of a symmetry
altering the overall sign, the related observable vanishes. Two such examples are shown in Fig.~\ref{fig:polygon}.
Consider, for instance, the invariant
\be
\label{polygon}
\sum_{a_1\dots a_n} M_{a_1a_2}M_{a_2a_3}\dots M_{a_{n}a_1}=\text{tr}\,\mathcal{M}^n
\ee
where, in contrast to $M$ defined in \eqref{momentum}, $\mathcal{M}=(M_{ab})$
is the angular momentum tensor treated as a matrix.
It is described by an $n$-sided polygon as shown in Fig.~\ref{fig:polygon}.
For odd values of $n$, they vanish due to the antisymmetry with respect to the inversion
$M_{ab}\to M_{ba}$ of all arrows.
For even values of $n$, they  correspond to the Casimir invariants
of $SO(N^2)$. As can be easily verified, the crossing relation \eqref{crossing}
implies that
$$
\mathcal{M}^n=\frac{1}{2}\left(\text{tr}\,\mathcal{M}^2\right)\mathcal{M}^{n-2}=\ical \mathcal{M}^{n-2}.
$$
Hence, the invariant \eqref{polygon} is just a power of the spherical
Hamiltonian  (see Fig.~\ref{fig:polygon}a),
$$
\text{tr}\,\mathcal{M}^n=\ical^{n/2}.
$$

Using the relation \eqref{d-un-sun}, which expresses the  $U(N)$ invariant tensors in terms of $SU(N)$ ones,
one can extend to the spherical invariants the relation  between the invariants of the conformal
mechanics with and without center of mass~\eqref{int-un-sun}.
For the spherical Hamiltonians this relation is given by   \eqref{ical-red}.
The general case can be treated using the split sum relation
\be
\sum_{a_1,a_2=0}^{N^2-1}d^{(k_1)}_{\dots a_1 \dots}d^{(k_2)}_{\dots a_2 \dots}M_{a_1a_2}
=\sum_{b_1,b_2=1}^{N^2-1}d^{(k_1)}_{\dots b_1 \dots}d^{(k_2)}_{\dots b_2 \dots}M_{b_1b_2}
+\sfrac{1}{\sqrt{N}}\sum_{b=1}^{N^2-1}\left(d^{(k_1)}_{\dots b \dots}d^{(k_2-1)}_{\dots \dots}
-d^{(k_1-1)}_{\dots  \dots}d^{(k_2)}_{\dots b \dots}
\right)M_{b0},
\ee
where $k_1,k_2$ denotes the order of the two invariant tensors.
As a result, a spherical invariant decomposes into $3^{\sum k}$ parts and can be combined
to a polynomial in $M_{0b}$. Its free term is just the version of the original integral
with excluded center of mass.

\begin{figure}
\begin{minipage}[b]{0.5\textwidth}
\begin{pspicture}(-0.8,-1.2)(2.6,0.2)
\rput(-0.6,-0.3){\large$a)$}
\multiplet{4}(1.6,-0.6){A}
\multiplet{4}(0,-0.6){B}
\conn{B4}{A1}
\ncarc[arcangle=30]{B3}{A2}
\ncarc[arcangle=40]{B2}{A3}
\ncarc[arcangle=45]{B1}{A4}
\end{pspicture}
\hspace{1cm}
\begin{pspicture}(-1.6,-1.2)(2,0.2)
\rput(-1.4,-0.3){\large$b)$}
\multiplet{4}(0,-0.6){A}
\dotnode[linecolor=blue](-1,-0.6){B}
\dotnode[linecolor=blue](1.9,-0.6){C}
\conn{A1}{B}\ncarc[arcangle=-45]{A2}{B}
\conn{A4}{C}\ncarc[arcangle=45]{A3}{C}
\end{pspicture}
\hspace{1cm}
\begin{pspicture}(-1.4,-1.2)(2,0.2)
\rput(-1.2,-0.3){\large$c)$}
\dotnode[linecolor=blue](0.5,0){A}
\multiplet{3}(-0.6,-0.6){B}
\multiplet{3}(0.9,-0.6){C}
\conn{B1}{A}
\conn{C3}{A}
\ncarc[arcangle=30]{B2}{C2}
\ncline{B3}{C1}
\end{pspicture}
\caption{\label{fig:four-order}  The three independent fourth-order invariants.}
\end{minipage}
\begin{minipage}[b]{0.4\textwidth}
\begin{pspicture}(-1.6,-1.2)(2,0.2)
\dotnode[linecolor=blue](-1.4,-0.6){A}
\multiplet{5}(-0.6,-0.6){B}
\multiplet{3}(1.3,-0.6){C}
\conn{B1}{A}
\ncarc[arcangle=-45]{B2}{A}
\ncarc[arcangle=45]{B3}{C3}
\ncarc[arcangle=35]{B4}{C2}
\conn{B5}{C1}
\end{pspicture}
\caption{\label{fig:five-order} The unique fifth-order invariant.}
\end{minipage}
\end{figure}

\section{Independent invariants and free-particle limit}
\noindent
>From the above observation, hereafter, we consider the spherical mechanics without center of mass.
It is a superintegrable system on the $(N-2)$-dimensional sphere with $2N-5$ functionally independent
constants of motion.
Apart from the spherical Hamiltonian itself (see Fig.~\ref{fig:examples-1}c),
there are no invariants of second order
in momenta (or angular momenta). There is no nontrivial third-order invariant,
but we have
three independent invariants of fourth order
and a single fifth-order invariant,
which are depicted respectively in Figs.~\ref{fig:four-order}a and \ref{fig:four-order}b.
Here, functional independence is understood in the $N\to\infty$ limit.
For fewer particles, additional algebraic relations restricting the number of independent
integrals appear.

\begin{figure}
\includegraphics[width=0.7\textwidth]{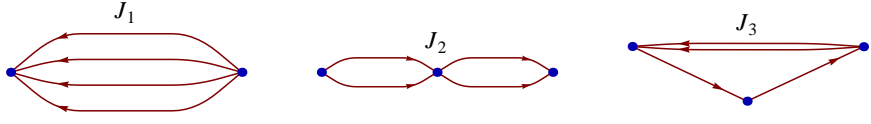}
\caption{The highest-order terms for the three independent fourth-order invariants. }
\label{fig:free-four-order}
\end{figure}

\begin{figure}
\includegraphics[width=0.7\textwidth]{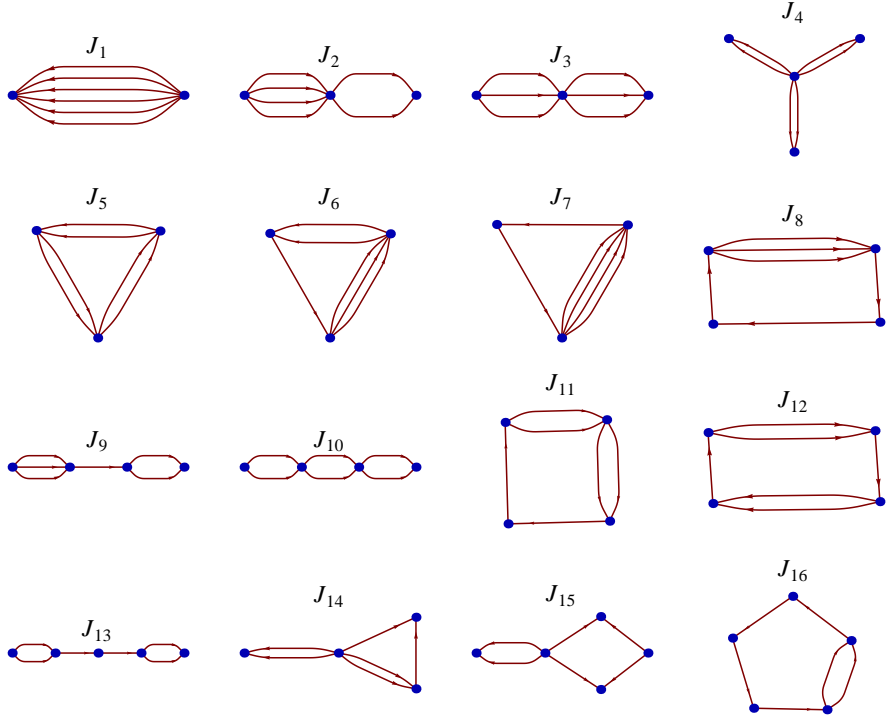}
\caption{The highest-order terms for all independent sixth-order invariants.}
\label{fig:free-six-order}
\end{figure}

There are sixteen sixth-order invariants.
In order to simplify the graphics and save space,
we collapse each leg into a single dot. For example, the fourth-order diagrams in Fig.~\ref{fig:four-order}
are equivalent to those depicted in  Fig.~\ref{fig:free-four-order}.
Then all independent sixth-order invariants are shown in  Fig.~\ref{fig:free-six-order}.
The others either vanish or are expressed in terms of second- and fourth-order
invariants. This result can be established by computer algebra
and applying the free-particle limit of the spherical mechanics.

In the $g\to0$ limit the constants of motion significantly simplify
like they do for the full Calogero model.
Recall that the reduction  of the matrix model to the Calogero system
preserves only the diagonal elements of the matrix of coordinates \eqref{redQ}.
In the free-particle limit, in addition, solely the diagonal elements of $P=L$ survive
according to \eqref{Liouville}.
Therefore, only the components $Q_k$, $P_k$ and $M_{kl}$
corresponding to the Cartan subalgebra survive in the free-particle limit.
Note that the restriction of the invariant tensor to the Cartan subalgebra are
totally symmetric.
The $g\to 0$ limit corresponds to the highest-order term in momenta:
\be
I_k(g=0)=\sum_{i_1\dots i_k=1}^{N}d'_{i_1\dots i_k}p_{i_1}\dots p_{i_k}=\sum_{i=1}^{N}p_i^k.
\ee
Here the usual coordinates of the Calogero model are used for simplicity. The related basis
is obtained by replacing $T_i\to E_{i+1\,i+1}$ in \eqref{cartan1}, \eqref{cartan2}, and
the decomposition of a diagonal element is $P=\sum_{i=1}^N p_iE_{ii}$.
In this basis, the indices of nonzero entries of the $U(N)$ invariant tensor coincide:
\be
\label{coincide}
d'_{i_1i_2\dots i_k}=\tr{E_{i_1i_1}\dots E_{i_ki_k}}=\delta_{i_1i_2}\delta_{i_1i_3}\dots\delta_{i_1i_k},
\qquad
1\le i_l\le N.
\ee

Coming back to the invariants of the spherical mechanics, we see
that the simplified graphs defined in Figs.~\ref{fig:free-four-order}
and \ref{fig:free-six-order} acquire a clear meaning in the free-particle limit:
a single point is labeled by a single index.
For example, the highest-order terms for the invariants \eqref{Ik-sp}
and \eqref{Ik'} are, respectively,
$$
\ical_k(g=0)=\sum_{i,j=1}^{N} M_{ij}^{2k},
\qquad
\ical'_k(g=0)=\sum_{i=1}^{N} (\mathcal{M}^2_{ii})^{k}.
$$
In Fig.~\ref{fig:free-six-order} the highest-order terms of all sixth-order invariants
are presented. For a large enough number of particles they are independent.

As an example, consider the two-dimensional spherical mechanics
$\ical^\red$ inherited from the four-particle Calogero model with
reduced center of mass. Apart from the Hamiltonian, it has two
independent invariants: one of them  can be chosen  to be of
fourth order in momenta, the other one of sixth order. The remaining
constants of motion are expressed in terms of them. In our recent
paper  \cite{hlns}, we have constructed such type of spherical
invariants starting from ones for the conformal mechanics. Now,
using computer algebra, we can express them in terms of the
diagrammatic invariants:
\begin{center}
\begin{pspicture}(-1,-0.8)(3,0.8)
\psset{linewidth=0.5pt}
%\showgrid
\rput[r](0.7,0){${\displaystyle J_2\sim\frac{3}{4}}$}
\rput[l](1.3,0){${\displaystyle -\;\frac12}$}
\dotnode[linecolor=blue](1,-16pt){1}
\dotnode[linecolor=blue](1,0){2}
\dotnode[linecolor=blue](1,16pt){3}
\ncarc[arcangle=35]{-}{1}{2}
\ncarc[arcangle=-35]{-}{1}{2}
\ncarc[arcangle=35]{-}{2}{3}
\ncarc[arcangle=-35]{-}{2}{3}
\multiput(2.1,0)(0.3,0){2}{
\dotnode[linecolor=blue](0,-12pt){up}
\dotnode[linecolor=blue](0,12pt){down}
\ncarc[arcangle=-25]{-}{up}{down}
\ncarc[arcangle=25]{-}{up}{down}
}
\rput[t](2.9,0){,}
\end{pspicture}
\hspace{2cm}
\begin{pspicture}(-1,-0.8)(5,0.8)
%\showgrid
\psset{linewidth=0.5pt}
\rput[r](0.7,0){${\displaystyle J_{3/2}^{1/2}=\frac{21}{6}}$}
\dotnode[linecolor=blue](1,-14pt){1}
\dotnode[linecolor=blue](1,14pt){2}
\ncarc[arcangle=15]{-}{1}{2}
\ncarc[arcangle=-15]{-}{1}{2}
\ncarc[arcangle=40]{-}{1}{2}
\ncarc[arcangle=-40]{-}{1}{2}
\ncarc[arcangle=80]{-}{1}{2}
\ncarc[arcangle=-80]{-}{1}{2}
\rput[l](1.5,0){${\displaystyle -\;\frac32}$}
\dotnode[linecolor=blue](2.4,-16pt){a}
\dotnode[linecolor=blue](2.4,0){b}
\dotnode[linecolor=blue](2.4,16pt){c}
\ncarc[arcangle=20]{-}{a}{b}
\ncarc[arcangle=-20]{-}{a}{b}
\ncarc[arcangle=30]{-}{b}{c}
\ncarc[arcangle=-30]{-}{b}{c}
\ncarc[arcangle=55]{-}{b}{a}
\ncarc[arcangle=-55]{-}{b}{a}
\rput[l](2.8,0){${\displaystyle +\;\frac{3}{64}}$}
\multiput(3.8,0)(0.3,0){3}{
\dotnode[linecolor=blue](0,-12pt){up}
\dotnode[linecolor=blue](0,12pt){down}
\ncarc[arcangle=-25]{-}{up}{down}
\ncarc[arcangle=25]{-}{up}{down}
}
\rput[t](4.8,0){.}
\end{pspicture}
\end{center}
Note that the double-dot leg in the last terms of both expressions corresponds to the
Hamiltonian.

\section{Relation with the valence-bond basis}
\noindent
In the previous sections, we have constructed the constants of
motion of the spherical mechanics by combining the conformal algebra
invariants $M_{ab}$ with the help of unitary invariant tensors.
Alternatively, one can obtain the same constants by proceeding in the opposite
order, namely by combining the unitarily invariant multiplets
of the conformal algebra into conformal singlets.
This method is  dual to the previous one.
Although the results are similar, the first approach is simpler for
the description and for calculations while the second one interprets
the constants of motion in terms of the valence-bond basis for
spin singlets~\cite{TL71}.

First, we consider the observables of the matrix mechanics \eqref{matrix}, which
are formed by the traces of the strings formed by a product of $P$ and
$Q$ matrices:
\be
\label{A-string}
O^{\sigma_1\dots\sigma_n}=
\sum_{a_1,\dots a_n} d_{a_1\dots a_n}A^{\sigma_1}_{a_1}\dots A^{\sigma_n}_{a_n}
=\tr{A^{\sigma_1}\dots A^{\sigma_n}},
\qquad
\sigma_i=\pm,\quad A^+=P, \quad A^-=Q.
\ee
Here we use $A^{\pm}$ instead of the dynamical variables for later convenience.
The canonical Poisson brackets now read
\be
\label{brackets}
\{A_{ij}^\sigma,A_{j'i'}^{\sigma'}\}=\epsilon_{\sigma\sigma'}\delta_{ii'}\delta_{jj'},
\ee
where  $\epsilon_{\sigma\sigma'}$ is the antisymmetric tensor
with $\epsilon_{+-}=1$.
The quantities \eqref{A-string} are $SU(N)$ invariant. Due to the above brackets,
they form the linear Poisson algebra \cite{etingof}
\be
\label{A-pois}
\left\{ O^{\sigma_1\dots\sigma_n},O^{\sigma'_1\dots\sigma'_m} \right\}
=\sum_{i,j}\epsilon_{\sigma_i\sigma'_j}O^{\sigma_{i+1}\dots \sigma_{i-1}\sigma'_{j+1}\dots \sigma'_{j-1}}.
\ee
This relation follows also from the completeness relation among the $U(N)$ invariant tensors \eqref{d-comp1}.
Here the cyclic ordering in the last trace is implied.  Note that, in the quantum case, polynomial corrections
appear due to the ordering issue between the matrix elements \cite{matsuo06}.

The angular momentum components can be expressed in terms of the newly defined
matrices as
\be
\label{M-eps}
M_{ab}=\sum_{\sigma_1,\sigma_2}\epsilon_{\sigma_1\sigma_2}A^{\sigma_1}_aA^{\sigma_2}_b.
\ee
The invariants of spherical mechanics may be expressed
in terms of \eqref{A-string} and $\epsilon_{\sigma\sigma'}$.
Namely, the $i$th dot of the related diagram (see, for instance,
Figs.~\ref{fig:examples-1} and \ref{fig:examples-2})
is marked now by $\sigma_i$, while the legs and bonds
are associated with
$O^{\sigma_1\dots\sigma_n}$ and $\epsilon_{\sigma_1\sigma_2}$
correspondingly, as is shown in Fig.~\ref{fig:epsilon}. The sum is taken over all $\sigma_i$.
This is a ``dual" interpretation of the diagram. It
can be obtained by substituting \eqref{M-eps} into
expressions for the invariants, like \eqref{Ik-sp} or \eqref{Ik'},
with the subsequent use of \eqref{A-string} and \eqref{inv-tensor}.
Commutation relations between the invariants, analogous to  \eqref{Inm} and \eqref{Inm-red},
can then be derived using the necklace relations \eqref{A-pois}.

\begin{figure}
\psset{arrows=->,dotsize=2pt}
\psset{linecolor=brown!70!black}
\begin{center}
\begin{pspicture}(0,0)(3,0.5)
\put(0.2,0){$\epsilon_{\sigma\sigma'}=$}
\pnode(1.5,2pt){a}
\pnode(2.8,2pt){b}
\ncline{a}{b}
\uput[u](1.5,1pt){\small $\sigma$}
\uput[u](2.8,1pt){\small $\sigma'$}
\end{pspicture}
\begin{pspicture}(-2,0)(2.0,0.8)
\put(-0.8,0){$(A^{\sigma_1}\dots A^{\sigma_5})=$}
\multido{\in=1+1,\nx=2+0.4}{5}{
\psdots[linecolor=blue!80!black](\nx,2pt)
\setcounter{count}{\the\multidocount}
\uput[u](\nx,2pt){\small $\sigma_\arabic{count}$}
}
\psellipse[linecolor=lightgray](2.8,2pt)(1.2,0.15)
\end{pspicture}
\end{center}
\caption{\label{fig:epsilon}
The ``dual" graphical representation.}
\end{figure}

Of course, only a finite amount of the described observables are independent.
Moreover, upon the reduction to the Calogero system by the gauge
fixing \eqref{redQ} and \eqref{redPQ}, the commutation
relations \eqref{redPQ} establish some order in a string of
matrices $P$ and $Q$.
Most natural is either the normal ordering or the Weyl (symmetrized) ordering \cite{gonera00,wadati93,avan95},
which are defined, respectively, by
\begin{align}
\label{Inm2}
&I_{kl}=\tr{P^kQ^l},
\\
&I^\text{sym}_{kl}=\text{Sym}(P^{k}Q^l)=
\binom{k+l}{l}^{-1}\oint\frac{dz}{2i\pi z^{l+1}}\text{tr}(P+zQ)^{k+l}.
\label{Isym}
\end{align}
The symmetrized traces \eqref{Isym} with $k+l=2s$ form a $(2s+1)$-dimensional
(nonunitary) $sl(2,R)$-representation of conformal spin
$s$ with the highest weight vector given by the Liouville integral
$I_{2s}$ of the Calogero model
\eqref{Ik}.
Using the usual notation of representation theory, they read
\begin{align}
\label{sm}
&I^\text{sym}_{s+m\,s-m}
=O^{(\sigma_1\dots\sigma_{2s})}
=\binom{2s}{s-m}^{-\frac12}\spin{s}{m},
\qquad
\text{where}
\quad
(\sigma_1\dots\sigma_{2s})=(\underbrace{+\dots+}_{s+m}\underbrace{-\dots-}_{s-m})
\end{align}
and $2m=k-l$. The symmetrization is performed over the bracketed indices.
The action of the conformal generators \eqref{gen-sl2} can be verified using the canonical
brackets \eqref{brackets}.

The quantities $I_{k0}=I_k$ and $I_{k1}$ coincide with their symmetrized counterparts and
form a closed Lie algebra. Their quadratic combinations $I_kI_{l1}-I_lI_{k1}$
lead to the  additional integrals of the Calogero system \cite{woj83}.
However, the whole set of observables \eqref{Inm2} or \eqref{Isym} does not form a closed Poisson algebra.
Indeed, in the case of normal ordering, according to  \eqref{A-pois}
we have the following decomposition of the Poisson-bracket structure constants:
$$
\{I_{kl},I_{k'l'}\}=(kl'-k'l)I_{k+k'-1\,l+l'-1}
+g\cdot\{\text{lower-length strings}\}.
$$
Here the strings with no more than $l+l'-2$  copies of $Q$ and $k+k'-2$  copies of $P$  appear after their rearrangement
according to  the normal ordering.
Apart from $I_{ij}$, these ordering terms contain polynomials in $\langle v|P^kQ^l|v\rangle$,
which together with the $I_{ij}$ form a closed nonlinear algebra \cite{avan95}.

A similar structure occurs for the brackets between Weyl-ordered observables,
\be
\{ \spin{s_1}{m_1},\spin{s_2}{m_2}\}=
2(s_2m_1-s_1m_2)\spin{s_1+s_2-1}{m_1+m_2}
+g\cdot\{\text{ordering terms}\},
\ee
where the lower-order terms, in general, have no symmetrized form
like \eqref{Isym} or \eqref{sm}.
Since the $sl(2,R)$ algebra acts additively on the Poisson brackets due to the Jacobi identity,
we get some kind of angular momentum sum rule for the $sl(2,R)$ representations of conformal
spins $s_1$ and $s_2$.
Here we have a kind of non-associative wedge product, where only the symmetric terms survive:
\be
\label{wedge}
(s_1)\wedge (s_2)=(s_1+s_2-1)\oplus (s_1+s_2-3)\oplus \ldots
=\bigoplus_{k=0}^{\left[\frac{s_1+s_2-1}2\right]} (s_1+s_2-1-2k).
\ee
In fact, the above properties of the Weyl-ordered quantities follow from the symmetrized analogue of the necklace relation,
\be
\label{A-sym-pois}
\left\{ O^{(\sigma_1\dots\sigma_{2s})},O^{(\sigma'_1\dots\sigma'_{2s'})} \right\}
=\sum_{\sigma_i,\sigma'_j}\epsilon_{\sigma_i\sigma'_j}O^{(\sigma_1\dots \widehat{\sigma_i} \dots \sigma_{2s})(\sigma'_1\dots\widehat{\sigma'_j}\dots \sigma'_{2s'})}.
\ee
The right part of this equation has the structure of a tensor product of two $sl(2,R)$ multiplets with
conformal spins $s-\sfrac12$ and $s'-\sfrac12$. Its contraction  by
$\epsilon_{\sigma_1\sigma'_1}\dots\epsilon_{\sigma_k\sigma'_k}$ with subsequent symmetrization over the
remaining spins
projects into the spin $s+s'-k-1$ representation. Since the left part of \eqref{A-sym-pois}
is antisymmetric under the exchange of the two spin sets $\{\sigma\}$ and $\{\sigma'\}$, only even values
of $k$ survive, which proves \eqref{wedge}.
 Using \eqref{sm} and the Clebsch-Gordan decomposition,
one can derive from \eqref{A-sym-pois} the following brackets between
the usual spin projection states:
\be
\begin{split}
\label{psi-pois}
\left\{ \psi_{sm},\psi_{s'm'} \right\}
&=\!\!\sum_{k=0}^{\left[\frac{s_1+s_2-1}2\right]}\left(\sqrt{(s{+}m)(s'{-}m')}C_{s-\frac12 m-\frac12,s'-\frac12 m'+\frac12}^{s+s'-2k-1\, m+m'}
-\sqrt{(s{-}m)(s'{+}m')}C_{s-\frac12 m+\frac12,s'-\frac12 m'-\frac12}^{s+s'-2k-1\, m+m'}\right)
\\
&\times2\sqrt{ss'}\, \psi_{s+s'-2k-1\, m+m'}.
\end{split}
\ee
One may replace all strings \eqref{A-string} in the expression for the spherical invariant
by their symmetrized counterparts \eqref{sm}. In the dual description of the
previous section, this substitution is equivalent to the use of symmetrized
$SU(N)$-invariant tensors
$$
d^\text{sym}_{a_1\dots a_n}=\frac{1}{n!}\sum_{p\in S_n} d_{a_{p_1}\dots a_{p_n}}.
$$
Of course, the new set of invariants differs from the old one. Although they
appear to be more cumbersome, the related diagrams are simpler.
Due to the symmetry,
all dots which form a multiplet are equivalent in a particular string.
So, it is more natural to use a single dot for a multiplet instead of a leg
as in the previous section, where the free-particle limit has been discussed.
In contrast to that case, however,
the dots (see, for example, Figs.~\ref{fig:free-four-order}
and \ref{fig:free-six-order})
must be labeled by multiple indices as before.

The relation \eqref{crossing} depicted in Fig.~\ref{fig:crossing}
is the well-known valence-bond crossing relation for four spin-1/2 singlet states.
In the theory of spin systems, the described states are known as a valence-bond basis
for spin-singlet states.
It is overcomplete. A true basis is formed by the Temperley-Lieb noncrossing states:
the multiplets are positioned along a single line or circle, and any bond distribution is allowed if it
respects the spin of the multiplet and avoids the crossing. The Temperley-Lieb basis
is nonorthogonal, even for $su(2)$ spins. All states presented in Figs.~\ref{fig:free-four-order} and
\ref{fig:free-six-order} are elements of the Temperley-Lieb basis.
After the $SU(N)$ reduction the constructed invariants form a
functional representation for $SL(2,R)$ singlet states, like the
simplest one considered in \cite{mam08} for the $SU(2)$ $s=1/2$ spins.

%\newpage

\noindent
{\bf Acknowledgments.} \\
T.H.\ and A.N.\ are grateful for hospitality at Leibniz Universit\"at Hannover,
where this study was initiated and where an essential part of the work has been
completed. This work was partially supported by the Volkswagen Foundation grant
I/84 496 and by the grants SCS 11-1c258 and SCS-BFBR 11AB-001
of the Armenian State Committee of Science.

%\newpage

\appendix
\section{Generators of the $U(N)$ group}
\noindent
Here we present an orthogonal basis of Hermitian matrices, which respects the
decomposition   $u(N)=u(1)\oplus su(N)$. Let
\be
\label{cartan1}
T_0=\frac{1}{\sqrt{N}} \mathbf{1}
\ee
be the $u(1)$ phase generator, and the remaining $N-1$ generators
span the $su(N)$ subalgebra:
\be
\label{cartan2}
T_k=\frac{1}{\sqrt{k(k+1)}}\left(\sum_{i=1}^k E_{ii}-kE_{kk}\right),
\qquad 1\le k\le N-1.
\ee
Here, $E_{ij}$ are the matrices with vanishing entries except for
one in the $i$th row and $j$th column.
The remaining $T_a$ are given by the off-diagonal matrices
\be
\label{borel}
\frac{1}{\sqrt{2}}(E_{jk}+E_{kj}), \qquad \frac{i}{\sqrt{2}}(E_{jk}-E_{kj}), \qquad j>k.
\ee
Together with the orthogonality condition \eqref{Ta}, the basic matrices obey
\be
\label{a0}
\text{tr}\, T_a=\delta_{a0}.
\ee
In addition, all generators obey the
$u(N)$ completeness relation
\be
\label{comp1}
\sum_{a=0}^{N-1}T^a_{ij}T^a_{kl}=\delta_{il}\delta_{jk}.
\ee
The corresponding relation for the $su(N)$ generators is
\be
\label{comp2}
\sum_{a=1}^{N-1}T^a_{ij}T^a_{kl}=\delta_{il}\delta_{jk}-\frac{1}{N}\delta_{ij}\delta_{kl}.
\ee

After the $su(N)$ reduction, the diagonal entries of the reduced matrices
\eqref{redQ} and \eqref{redP} define the coordinates and momenta of the
Calogero system \eqref{Calogero}. Using  the definitions
\eqref{cartan1} and \eqref{cartan2}, their relation with the new coordinates
$Q_i=\tr{QT_i}$ and $P_i=\tr{PT_i}$ can be derived \cite{jacobi02,cuboct}:
\be
\label{jacobi}
\begin{aligned}
 Q_k=\sfrac{1}{\sqrt{k(k+1)}}(q_1+\dots +q_k-kq_{k+1}), \quad 1\le k\le N-1,
\qquad
Q_0=\sfrac{1}{\sqrt{N}}(q_1+\dots +q_N),
\\
P_k=\sfrac{1}{\sqrt{k(k+1)}}(p_1+\dots +p_k-kp_{k+1}), \quad 1\le k\le N-1,
\qquad
P_0=\sfrac{1}{\sqrt{N}}(p_1+\dots +p_N).
\end{aligned}
\ee
These are the usual Jacobi
coordinates, which are used in scattering theory in order to eliminate the center of
mass \cite{reed78}. Here it corresponds to the $U(1)$ reduction \cite{poly99}.
The center of mass is excluded simply by imposing $Q_0=P_0=0$.
Note that $q_i=\tr{QE_{ii}}$ and $p_i=\tr{PE_{ii}}$. Since both bases $E_{ii}$ and $T_i$
are orthonormal, the transformation \eqref{jacobi} is orthogonal. Therefore, the kinetic therm
of the original Calogero model remains unchanged,
\be
\label{Calogero-red}
H^\red=\frac{1}{2}\sum_{i=1}^{N-1} P_i^2 + \sum_{1\le i<j\le N-1}\frac{g^2}{(\alpha_{ij}Q)^2},
\ee
where $\alpha_{ij}=E_{ii}-E_{jj}$ are the roots of the $su(N)$ algebra.


\begin{thebibliography}{99}

\bibitem{calogero69}
F.~Calogero, J.\ Math.\ Phys. {\bf 10} (1969) 2191;
{\sl ibid.} {\bf 12} (1971) 419.

\bibitem{trig-Cal}
B.~Sutherland, Phys.\ Rev.\ {\bf A4} (1971) 2019;
Phys. Rev. {\bf A5}(1972) 1372.

\bibitem{spin-Cal}
J.~Gibbons and T.~Hermsen, Physica {\bf 11D}(1984) 337;

S.~Wojciechowski, Phys.\ Lett.\ A {\bf 111} (1985) 101.

\bibitem{super-Cal}
D.Z.~Freedman and P.F.~Mende,
%``AN EXACTLY SOLVABLE N PARTICLE SYSTEM IN SUPERSYMMETRIC QUANTUM
%MECHANICS,''
Nucl.\ Phys.\  B {\bf 344} (1990) 317.
%%CITATION = NUPHA,B344,317;%%

S.~Fedoruk, E.~Ivanov and O.~Lechtenfeld,
%``Supersymmetric Calogero models by gauging,''
Phys.\ Rev.\  D {\bf 79} (2009) 105015.
%[arXiv:0812.4276 [hep-th]].
%%CITATION = PHRVA,D79,105015;%%

\bibitem{algebra}
J.~Wolfes,
%``On the three-body linear problem with three-body interaction,''
J.\ Math.\ Phys.\  {\bf 15}, 1420 (1974).
%%CITATION = JMAPA,15,1420;%%

F.~Calogero and C.~Marchioro,
% ``Exact solution of a one-dimensional three-body scattering problem with
%two-body and/or three-body inverse-square potentials,''
J.\ Math.\ Phys.\  {\bf 15}, 1425 (1974).
%%CITATION = JMAPA,15,1425;%%

M.A.~Olshanetsky and A.M.~Perelomov,
% ``Quantum Completely Integrable Systems Connected With Semisimple Lie
%Algebras,''
Lett.\ Math.\ Phys.\  {\bf 2}, 7 (1977).
%%CITATION = LMPHD,2,7;%%

\bibitem{hlkn}
T.~Hakobyan, S.~Krivonos, O.~Lechtenfeld and A.~Nersessian,
%``Hidden symmetries of integrable conformal mechanical systems,''
Phys.\ Lett.\  A {\bf 374} (2010) 801.
% [arXiv:0908.3290 [hep-th]].
%%CITATION = PHLTA,A374,801;%%

\bibitem{woj83}
S.~Wojciechowski, Phys.\ Lett.\ A {\bf 95} (1983) 279.

\bibitem{cuboct}
T.~Hakobyan, A.~Nersessian and V.~Yeghikyan,
%``Cuboctahedric Higgs oscillator from the Calogero model,''
J.\ Phys.\ A  {\bf 42} (2009) 205206.
%[arXiv:0808.0430 [math-ph]].
%%CITATION = JPAGB,A42,205206;%%

\bibitem{feigin}
M.V.~Feigin,
Theor.\ Math.\ Phys.\ {\bf 135} 497 (2003).

\bibitem{hlns}
T.~Hakobyan, O.~Lechtenfeld, A.~Nersessian and A.~Saghatelian,
%``Invariants of the spherical sector in conformal mechanics,''
J.\ Phys.\ A  {\bf 44} (2011) 055205.
% [arXiv:1008.2912 [hep-th]].
%%CITATION = JPAGB,A44,055205;%%
%\cite{Lechtenfeld:2010xu}

\bibitem{perelomov}
M.A.~Olshanetsky and A.M.~Perelomov,
%``Classical integrable finite dimensional systems related to Lie algebras,''
Phys.\ Rept.\  {\bf 71}, 313 (1981);
%%CITATION = PRPLC,71,313;%%
%``Quantum Integrable Systems Related To Lie Algebras,''
Phys.\ Rept.\  {\bf 94}, 313 (1983).
%%CITATION = PRPLC,94,313;%%

\bibitem{TL71}
H.N.V.~Temperley and E.H.~Lieb,
Proc.\ R.\ Soc.\ London A {\bf 322} (1971) 251.

\bibitem{moser}
J.~Moser, Adv.\ Math.\ {\bf 16} (1975) 197.

\bibitem{polychronakos}
A.P.~Polychronakos, J.\ Phys.\ A  {\bf 39} (2006) 12793.

\bibitem{gauge}
C.~Gonera, P.~Kosinski, and P.~Maslanka, Phys.\ Lett.\ A {\bf 283} (2001) 119.

\bibitem{poly99}
%Generalized Statistics in One Dimension
A.P.~Polychronakos, in \emph{Topological aspects of low dimensional
systems}, Les Houches {\bf 69} (1999) 415.

\bibitem{woj84}
S.~Wojciechowski, Phys.\ Lett.\ A {\bf 104} (1984) 189.

\bibitem{etingof}
P.~Etingof and X.~Ma,
\emph{Lecture notes on Cherednik algebras} (2010)
arXiv:1001.0432v4[math.RT].

\bibitem{matsuo06}
Y.~Hatsuda and Y.~Matsuo,
J.\ Phys.\ A \textbf{40} (2007) 1633.

\bibitem{wadati93}
K.~Hikami and M.~Wadati, J.\ Phys.\ Soc.\ J.\ {\bf 62} (1993) 4203;
Phys.\ Rev.\ Lett.\ {\bf 73} (1994) 1191.

\bibitem{avan95}
J.~Avan and E.~Billey,
Phys.\ Lett.\ A \textbf{198} (1995) 183.

\bibitem{gonera00}
T.~Brzezinski, C.~Gonera, P.~Kosinski and P.~Maslanka,
Phys.\ Lett.\ A {\bf 268} (2000) 178.

\bibitem{mam08}
M.~Mambrini,
Phys.\ Rev.\ B \textbf{77} (2008) 134430.

\bibitem{jacobi02}
M.V.~Ioffe and A.I.~Neelov, J.\ Phys.\ A {\bf 35}, (2002) 7613.

\bibitem{reed78}
M. Reed and B. Simon,  \emph{Methods of modern mathematical physics}, vol III
(New York: Academic) 1978.

\end{thebibliography}
\end{document}